\documentclass{article}
\usepackage{scrextend}
\usepackage{geometry}
\usepackage[english]{babel}
\usepackage{amssymb}
\usepackage{tcolorbox}
\usepackage{amsmath}
\usepackage{amsthm}
\usepackage[normalem]{ulem}
\usepackage{float}
\useunder{\uline}{\ul}{}

\usepackage{xcolor}
\usepackage{amsmath}
\usepackage{amsfonts}
\usepackage{bbm}
\usepackage{algorithm}
\usepackage[noend]{algorithmic}
\usepackage{subcaption}
\usepackage{caption}
\usepackage{subcaption}
\newcounter{mechanism}

\newtheorem{theorem}{Theorem}
\theoremstyle{definition}
\newtheorem{definition}{Definition}
\newtheorem{Proposition}{Proposition}

\newtheorem{lemma}[theorem]{Lemma}
\newtheorem{claim}{Claim}
\usepackage{etoolbox}

\preto\equation{\par\nobreak\small\noindent}
\preto\align{\par\nobreak\small\noindent}

\title{Designing Truthful Contextual Multi-Armed Bandits based Sponsored Search Auctions}
\date{}
\author{Kumar Abhishek \\\small IIIT Hyderabad, India.\\\small kumar.abhishek@research.iiit.ac.in 
\and Shweta Jain\\\small IIT Ropar, India. \\\small shwetajain@iitrpr.ac.in
\and Sujit Gujar\\\small IIIT Hyderabad, India.\\\small sujit.gujar@iiit.ac.in}
\usepackage[utf8]{inputenc}
\begin{document}
\maketitle
\begin{abstract}
For sponsored search auctions, we consider contextual multi-armed bandit problem in the presence of strategic agents. In this setting, at each round, an advertising platform (center) runs an auction to select the best-suited ads relevant to the query posted by the user. It is in the best interest of the center to select an ad that has a high expected value (i.e., probability of getting a click $\times$ value it derives from a click of the ad). The probability of getting a click (CTR) is unknown to the center and depends on the user's profile (context) posting the query. Further, the value derived for a click is the private information to the advertiser and thus needs to be elicited truthfully. The existing solution in this setting is not practical as it suffers from very high regret ($O(T^{\frac{2}{3}})$). 

Towards designing practically useful mechanisms, we first design an elimination-based algorithm \emph{ELinUCB-SB} that is ex-post monotone, which is a sufficient condition for truthfulness. Thus, \emph{ELinUCB-SB} can be naturally extended to ex-post incentive compatible and ex-post individually rational mechanism \emph{M-ELinUCB-SB}. We show via experiments that the proposed mechanisms outperform the existing mechanism in this setting. Theoretically, however, the mechanism may incur linear regret in some instances, which may not occur frequently. To have a theoretically stronger mechanism for regret, we propose a \emph{SupLinUCB}-based allocation rule \emph{SupLinUCB-S}. With the help of \emph{SupLinUCB-S}, we design a mechanism \emph{M-SupLinUCB-S}, which is ex-post incentive compatible and ex-post individually rational. We prove that it has regret $O(n^2 \sqrt{dT \log T})$ as against $O(n\sqrt{dT \log T)}$ for non-strategic settings; $O(n)$ is price of truthfulness. 
We demonstrate the efficacy of our mechanisms via simulation and establish superior performance than the existing literature.
   

\end{abstract}

\section{Introduction}
Internet advertising is one of the booming and rapidly increasing industry with revenue volume in billions of dollars \cite{iab}. The majority of the revenue generated by search engines like Google, Yahoo, and Bing comes from advertisements displayed on their platform. Typically, for any search query by a user, a search engine/the advertising platform, henceforth a \emph{center}, displays ads along with the relevant results via an auction mechanism known as \emph{sponsored search auction} (SSA). The fundamental difference between traditional advertising and Internet advertising is the payment model. In the former, the advertisers (henceforth \emph{agents}), pay based on the number of impressions whereas in latter, the agents pay only if their ad receives a click.
Thus, the probability of an ad getting clicked, referred to as \emph{click-through rate} (CTR), plays a crucial role in SSA. The CTR of an ad is unknown to the center, but it can learn CTRs by displaying the ad repeatedly over a period of time. Each agent also has a private valuation for its ad, which represents its willingness to pay for a click. This valuation needs to be elicited from the agents truthfully. 
\par 
In the absence of contexts, when the agents report their true valuations, we can model the problem as a Multi-Armed Bandit (MAB) problem \cite{lai85} with agents playing the role of the arms. As the agents (arms) are strategic, they may misreport their valuations to maximize their utility. To elicit truthful bids from the agents, researchers have used Mechanism Design \cite{aggarwal06,Nisan07}. However, such mechanisms are oblivious to the learning requirements and fail to avoid manipulations by the agents when learning is involved. In such cases, the researchers have modeled this problem as a MAB mechanism \cite{Bab09,Nikh09,Gatti12,jain18}. The authors designed \emph{ex-post truthful} mechanisms wherein the agents are not able to manipulate even when the random clicks are known to them.   
\par 
Typically an individual user tends to click some specific ads more often than the other ads, which depends upon the profile of an individual user of the platform. In this work, we leverage this fact and use the profile of the user as features (for context) to personalize the ads to increase the number of clicks and hence, the \emph{social welfare}. When the CTRs of the ads depend on the specific context at a particular round, we can model the problem as a \emph{Contextual MAB} (ConMAB) problem \cite{Auer03,Lang07,Li10,Abbasi11}. However, a naive implementation of ConMAB is not adequate in the presence of strategic agents.
\par
To the best of our knowledge, contextual information in SSA is considered only in \cite{Gatti12}. The authors proposed a novel, theoretically sound, deterministic, exploration-separated mechanism that offers strong game-theoretic properties. However, it faces multiple practical challenges: (i) it incurs high cost of learning (\emph{regret}), (ii) the center needs to know the number of rounds for which it needs to execute SSA, and (iii) the initial rounds being free, a malicious agent may drop off after free rounds; in some cases, all the rounds could be free.
\subsection*{Contributions}
In the presence of strategic agents, random context-arrivals, and stochastic clicks, our goal is to design a non-exploration-separated, ex-post truthful mechanism that (i) learns CTRs efficiently (minimizes regret), (ii) may not need prior knowledge of $T$, and (iii) does not have free rounds. We leverage popular algorithms \emph{LinUCB} \cite{Li10} and \emph{SupLinUCB} \cite{Chu11} that perform well in estimating CTRs in the contextual setting to build our randomized mechanisms to avoid manipulations by strategic agents. In particular, our contributions are:
\begin{itemize}
     \item We adapt \emph{LinUCB} to design an ex-post monotone allocation rule \emph{ELinUCB-S} for a single-slot SSA (Theorem \ref{thm:linucb_bsln_mon}).
     We further optimize \emph{ELinUCB-S} by introducing batch level update to propose \emph{ELinUCB-SB} and using resampling procedure by \cite{Bab15}, we develop an ex-post truthful mechanism \emph{M-ELinUCB-SB}, which is also ex-post individually rational (Theorem \ref{thm:M-LinUCB-SB_EPIC}). Unlike existing ConMAB mechanism, \emph{M-ELinUCB-SB} does not need to know $T$.

    \item For stronger theoretical guarantees, we adapt \emph{SupLinUCB} to design an ex-post monotone allocation rule \emph{SupLinUCB-S} for a single-slot SSA (Theorem \ref{thm:M-SupLinUCB_EPIC}). We prove that \emph{SupLinUCB-S} has regret $O(n^2\sqrt{dT \log T})$ (Theorem \ref{thm:suplinucb_reg}) as against $O(n\sqrt{dT \log T})$ for the non-strategic settings; we attribute $O(n)$ as price of truthfulness. Using resampling procedure, we develop \emph{M-SupLinUCB-S} which is ex-post truthful and ex-post individually rational. \emph{M-SupLinUCB-S}, however, needs to know $T$ upfront. 
    
     \item We study \emph{M-ELinUCB-SB} and \emph{M-SupLinUCB-S} with the existing mechanism \emph{M-Reg}, on simulated data and provide empirical analysis.
     Empirically, \emph{M-ELinUCB-SB} performs superior to \emph{M-SupLinUCB-S} by large factors for less than million rounds. 
 \end{itemize}

\section{Preliminaries}
\label{sec:prelims}
First, we define our model and notation. 

\subsection{Model and Notation}
\label{ssec:notation}
 There is a fixed set of agents $\mathcal{N} = \{1,2,\dots,n\}$, where each agent has exactly one ad to display and the center has one slot available for allocation. A contextual $n-$armed Multi-Armed Bandit (MAB) mechanism $\mathcal{M}$ proceeds in discrete rounds $t = 1,2, \ldots, T$. At each round $t$:
\begin{enumerate}
\item $\mathcal{M}$ observes a context $x_t \in [0, 1]^d$ which summarizes the profile of the user arriving at round $t$.
\item Based on the history, $h_t$, of allocations, observed clicks, and the context $x_t$, $\mathcal{M}$ chooses an agent $I_t \in \mathcal{N}$ to display it's ad. 
\item A click $r_{I_t}$ is observed which is $1$ if it gets clicked and $0$ otherwise. 
\item Mechanism $\mathcal{M}$ decides the positive payment $p_{I_t,t}$ to be made by the agent $I_t$ to the center. The payment by any other agent is $0$. 
\item Update $h_t = h_{t-1} \cup \{x_{t}, \{I_{t}\}, \{r_{I_t}\}\}$.
\item The mechanism then improves its arm-selection strategy with new observation ($x_{t}, \{I_t\}, \{r_{I_t}\}$). No feedback is received for the agents that are not selected.
\end{enumerate}
Each agent $i$ is thus characterized by two quantities: (i) private valuation $v_i \in [0,1]$, which represents the willingness to pay for the click received and is constant throughout the rounds, and (ii) click through rate (CTR) of its ad $\mu_i(x_t) \in [0, 1]$ which is an unknown parameter and is dependent on the context $x_t$. Each agent $i$ submits the valuation of getting a click on its ad as bid $b_i$. We assume that the bids are constant across the rounds (since $v_i$'s are constant). We assume that the CTR of an agent $i$ is linear in $d$-dimensional context $x_{t}$ with some unknown coefficient vector $\theta_i$ \cite{Li10}. Thus, the problem reduces to learning the $d-$dimensional vector $\theta_i$ for each agent $i$. The probability of getting a click on the ad of agent $i$  at any given round $t$ is given as:
$$\mathbbm{P}[r_{i,t}| x_{t}] = \mu_i(x_t) = \theta_i^\intercal x_{t}$$
Thus, the expected valuation of agent $i$ is $v_i \mu_i$. Let $b_{-i}$ be the bid vector of all the agents other than $i$ and $b$ denote the bid vector of all the agents. The utility of an agent $i$ in round $t$ with history $h_t$ is given as:
\begin{align*}
u_{i,t}(b_{i},b_{-i},x_t;h_t;v_i) = \mathbbm{1}\{I_t(b,x_t;h_t)=i \}r_{i,t}(v_i - p_{i,t}(b;h_t))    
\end{align*}
and the utility of center is given as: $$u^c_t(x_t) = \sum_{i=1}^{N} \{I_t(b,x_t;h_t)=i \}r_{i,t} p_{i,t}$$ 
In this work, our aim is to maximize social welfare  similar to \cite{Bab09,jain18}.
The social welfare at round $t$ is evaluated as sum of utilities of the agents and the center and is given as: $$sw_t(x_t) =  \sum_{i=1}^{N}\{I_t(b,x_t;h_t)=i \}r_{i,t} v_i.$$ 
When the CTRs are not known, the efficiency of any mechanism is measured by its rate of learning or regret. Thus, our goal reduces to design a mechanism $\mathcal{M}$ that minimizes the social welfare regret which is given as: 
\begin{equation}
    \mathbb{R}_T(\mathcal{M}) = \sum_{t=1}^T[\theta_{i^*_{t}}^Tx_t\cdot b_{i^*_{t}} - \theta_{I_t}^Tx_t\cdot b_{I_t}]
\end{equation}
Here, $i^*_{t}(x_t)$ denote the highest expected valuation (based on bids) i.e., $i^*_{t}(x_t) = argmax_k \{b_k \cdot (\theta_k^Tx_t)\}$.

In the following section, we define game theoretic properties relevant to this work.
\subsection{Game Theoretic Properties}
\label{ssec:gt_prop}
A mechanism $\mathcal{M}=(\mathcal{A}, \mathcal{P})$ (where $\mathcal{A}$ is the allocation rule and $\mathcal{P}$ is the payment rule) is ex-post truthful (formally called EPIC) if and only if $\mathcal{A}$ is ex-post monotone \cite{Myr81,arch01}.
That is, for all instances of possible click realizations and context-arrivals, by increasing its bid to $b_i^+>b_i$, agent $i$ should obtain at least same number of clicks at bid $b_i$ if not more. Formally,
 \begin{definition}
 \textbf{Ex-post monotonicity:} Let $\nu_i(b_i, t)$ denote total number of clicks on the ad of agent $i$ in first $t$ rounds. Then, $\mathcal{A}$ is ex-post monotone if for every possible sequence of context-arrivals and click realizations, for each agent $i \in \mathcal{N}, \forall t, \forall b_{-i}$ and two possible bids of $i$, $b_i^{+} \geq b_i$ we have $$\nu_i(b_i^{+}, t) \geq \nu_i(b_i, t).$$
 \end{definition}
\begin{definition}
 \textbf{EPIC:} A mechanism $\mathcal{M}=(\mathcal{A}, \mathcal{P})$ is said to be \emph{ex-post incentive compatible} ($EPIC$) if by misreporting the bid, no agent can gain its total utility more than that it would have obtained by bidding truthfully, i.e., $  \forall i, \forall b_{-i}, \forall v_i, \forall h_t,  \forall b_{i},$
 \begin{equation*}
    \sum_{t=1}^T u_{i,t}(v_i, b_{-i}, x_t;h_t;v_i)  \geq \sum_{t=1}^T u_{i,t}(b_{i}, b_{-i}, x_t;h_t;v_i).
 \end{equation*}
\end{definition}
EPIC implies even if an agent has observed all the contexts and all the click realizations, it is in the agent's best interest to report true valuation.  Note that, if a mechanism does any randomization, $u_{i,t}(\cdot)$ is replaced by $\mathbb{E} [u_{i,t}(\cdot)]$, where the expectation is taken w.r.t. randomization in the mechanism. Such a mechanism is still truthful for every realization of external randomness such as click realizations, context-arrivals. 

One alternative notion of IC which may seem suitable in our model is dynamic IC \cite{berg2010}. We would like the reader to note that in our model, the bids, as well as valuations, are constant throughout the rounds, and not dependent on any round $t$. Thus, private information and communication with the mechanism are not dynamic. Hence, a strong game-theoretic property of ex-post IC is more apt in our model.
\begin{definition}
\label{defn:epir}
 \textbf{EPIR:} A mechanism $\mathcal{M}=(\mathcal{A}, \mathcal{P})$ is said to be \emph{ex-post individually rational} ($EPIR$) if every agent has a non-negative utility with truthful bidding irrespective of the bids of other agents i.e., $\forall i, \forall x_t, \forall v_i, \forall t, \forall h_t$, $$u_{i,t}(v_i, b_{-i}, x_t;h_t;v_i) \geq 0.$$
\end{definition}
  
The authors in \cite{Bab15} have shown the power of randomization in designing truthful mechanisms by proposing a randomized context-free MAB mechanism that is ex-post truthful and has regret $O(\sqrt{T})$. Thus, by introducing randomness in the mechanism, they showed that it is possible to bypass the impossibility result in \cite{Bab09}, which states that any deterministic, truthful MAB mechanism has to be exploration-separated and hence must suffer a regret of $\Omega(T^{2/3})$. The main result of \cite{Bab15} involves designing the black box mechanism using the self-resampling procedure in Algorithm \ref{algo:resamp}, which provides ex-post truthful and IR mechanism if given an ex-post monotone allocation rule (Theorem \ref{thm:babaioff}). 
 
\begin{algorithm}[!ht] 
\begin{small}
\caption{Non-recursive self-resampling procedure \cite{Bab15}}
\label{algo:resamp}
\begin{algorithmic}[1]
\STATE \textbf{Input:} bid $b = \{b_1,\ldots, b_n\}$, parameter $\delta \in (0, 1)$
\STATE \textbf{Output:} modified bid $y =\{y_1,\ldots, y_n\}$, $\eta = (\eta_1, \ldots, \eta_n$)  
\STATE Independently for each agent $i \in \mathcal{N}$
\STATE \hspace{\algorithmicindent} Sample: $\epsilon_i$ uniformly at random from $[0, 1]$
\STATE \hspace{\algorithmicindent} \textbf{with probability} $1- \delta$ 
\STATE \hspace{\algorithmicindent} \hspace{\algorithmicindent} $\eta_i = 1$
\STATE \hspace{\algorithmicindent} \textbf{else} 
\STATE \hspace{\algorithmicindent} \hspace{\algorithmicindent} $\eta_i = \epsilon_{ i }^{ 1 / ( 1 - \delta ) }$
\STATE Construct the vector of modified bids $y = \left( y_{ 1 } , \dots , y_{ n } \right) , \text { where } y_{ i } = \eta_{ i } b_{ i }$
\end{algorithmic}
\end{small}
\end{algorithm}

\begin{theorem}
 \label{thm:babaioff}
 (Theorem 4.5, \cite{Bab15})
   Let $\mathcal{A}$ be ex-post monotone allocation rule. Applying the transformation in Algorithm \ref{algo:resamp} to $\mathcal{A}$ with parameter $\delta$, we obtain a mechanism $\mathcal{M}$ such that $\mathcal{M}$ is EPIC and EPIR. 
 \end{theorem}

There is no previous work done in designing the ex-post monotone allocation rule in the contextual setting. Hence, we address this problem and design two allocation rules, which are ex-post monotone, though it has different properties.

Concerning ConMAB mechanisms for SSA, two works are closely related to our work \cite{Gatti12} and \cite{Li10}. The former considers the strategic agents in ConMAB \cite{Gatti12} by proposing a mechanism that we will call \emph{M-Reg}. The mechanism is \emph{exploration-separated} \cite{Nikh09,Bab09} which is deterministic and induces EPIC property.
The regret achieved by this mechanism is quite high $O(T^{2/3})$ as compared to $O(\sqrt{T})$ regret in the traditional ConMAB problem. The latter introduces $LinUCB$ algorithm, which is particularly of interest to us. Hence we describe it below.

\subsection{\emph{LinUCB}}
\label{ssec:linUCB}
\emph{LinUCB} \cite{Li10} is a generic ConMAB algorithm that efficiently learns the CTR of an agent where the CTR model is linear in terms of context and the unknown parameters. 
The authors experimentally showed the efficacy of the algorithm in approximating the CTRs of news articles in news recommendation. Hence, we choose to adapt it to our setting.
\begin{algorithm}[!ht] 
\caption{\emph{LinUCB}}
\label{algo:linucb}
\begin{algorithmic}[1]
\STATE \textbf{Inputs:} $\alpha \in \mathbbm{R}_{+}$
\STATE \textbf{Initialization:} 
\FORALL {$i \in \mathcal{N}$}
    \STATE  $A_i \leftarrow I_d$ ($d$-dimensional identity matrix)
    \STATE $c_i \leftarrow 0_{d\times 1}$ ($d$-dimensional zero vector)
\ENDFOR
\FOR {$t = 1,2,3,\ldots,T$}
\STATE Observe context of user as $ x_{t} $
\FORALL {$i \in \mathcal{N}$}
\STATE $\hat{\theta}_{i} \leftarrow A_{i}^{-1}c_{i}$, $\mu_{i,t}^+ \leftarrow \hat{\theta}_{i}^\intercal x_t + \alpha \sqrt{x_{t}^\intercal A_{i}^{-1}x_{t}}$
\ENDFOR
\STATE $I_t = argmax_{i \in \mathcal{N}}\hspace{0.2em} \mu_{i,t}^+$, Observe $r_{I_t} \in \{0, 1\}$
\STATE $A_{I_t} \leftarrow A_{I_t} + x_{t}x_{t}^\intercal$, $c_{I_t} \leftarrow c_{I_t} + r_{I_t}x_{{t}}$
\ENDFOR
\end{algorithmic}
\end{algorithm}
\emph{LinUCB} is motivated by \emph{UCB} \cite{auer02} where upper confidence bound (UCB) is maintained for each agent. To capture the contextual information in ConMAB setting, \emph{LinUCB} uses $A_i $ and $c_i$ for each agent $i$, where $A_i$ summarizes the information about contexts and $c_i$ corresponding clicks. It maintains upper confidence bound (UCB) for each agent $i$ as $ \mu^+_{i,t} \leftarrow \hat{{\theta}}_{i}^{\top}x_{t}+\alpha \sqrt{x_{t}^{\top} A_{i}^{-1} x_{t}}$ where $\hat{\theta}_i \leftarrow A_{i}^{-1}c_{i}$ and $\alpha$ is learning parameter. At round $t$, the algorithm selects the agent $I_t$ with the highest UCB $\hat{\mu}_{i,t}$. The statistics for the selected agent $I_t$ is updated as $A_{I_t} \leftarrow A_{I_t} + x_{t}x_{t}^\intercal$, $c_{I_t} \leftarrow c_{I_t} + r_{I_t}x_{{t}}$, where $r_{I_t}$ is the indicator variable of receiving click. 

\emph{LinUCB} was originally designed to estimate CTRs of news articles and hence does not capture strategic manipulations. Motivated by \emph{LinUCB}, we build randomized EPIC mechanisms for SSA by developing an ex-post monotone allocation rule, \emph{ELinUCB-SB}, and using the resampling procedure (Algorithm \ref{algo:resamp}) to design a randomized EPIC and EPIR mechanism \emph{M-ElinUCB-SB} \cite{Bab15}. ELinUCB-SB has linear regret. We present it here as the key ideas to adapt \emph{LinUCB} to design truthful mechanisms are useful and carry forward when we design a more complicated mechanism, \emph{M-SupLinCUB-S}, based on \emph{SupLinUCB} by \cite{Chu11}.

\section{\emph{M-ELinUCB-SB}: Truthful ConMAB Mechanism 1}
\label{sec:our_approach2}

We first propose a single-slot allocation rule \emph{ELinUCB-S} based on \emph{LinUCB}. We next provide a further optimized algorithm \emph{ELinUCB-SB} that incorporates mini-batch learning, which makes the algorithm efficient both in terms of regret and computation. 

\subsection{\emph{ELinUCB-S}: LinUCB-Based Single-Slot SSA}
\begin{algorithm}[!ht] 
\caption{\emph{ELinUCB-S}: LinUCB-based allocation for single-slot SSA}
\label{algo:linucb_base}
\algsetup{indent=0.75em}
\begin{small}
\begin{algorithmic}[1]
\STATE \textbf{Inputs:} $n$, $\alpha \in \mathbbm{R}_{+}$, bid vector $b$ 
\STATE \textbf{Initialization:}  $S_{act} =$ $\mathcal{N}$
\FORALL {$i \in \mathcal{N}$}
    \STATE  $A_i \leftarrow I_d$ ($d$-dimensional identity matrix)
    \STATE $c_i \leftarrow 0_{d\times 1}$ ($d$-dimensional zero vector)
    \STATE $\mu_i^+ \leftarrow b_i; \mu_i^- \leftarrow 0$ 
\ENDFOR

\FOR {$t = 1,2,3,\ldots $}
    \STATE Observe context as $ x_{t} $
    \STATE $I_{t^{'}} \leftarrow 1+ (t \mod n)$
    \IF {$I_{t^{'}} \in S_{act}$}
    \STATE Allocate agent $I_{t^{'}}$, i.e., $I_t \leftarrow I_{t^{'}}$
    \STATE Observe click as $r_{I_t} \in \{0, 1\}$
    \STATE $A_{I_t} \leftarrow A_{I_t} + x_{t}x_{t}^\intercal$, $c_{I_t} \leftarrow c_{I_t} + r_{I_t}x_{{t}}$, $\hat{\theta}_{I_t} \leftarrow A_{I_t}^{-1}c_{I_t}$
    \STATE\{Update confidence bound\}
    \IF {$\mu_{I_t}^- < \mu_{I_t}^+$}
    \STATE $(\gamma_{I_t}^{-} , \gamma_{I_t}^{+}) \ \leftarrow b_{I_t}(\hat{\theta}_{I_t} x_{t} \mp \alpha \sqrt{x_{t}^\intercal A_{I_t}^{-1}x_{t}})$
    \IF {$\max(\mu_{I_t}^-, \gamma_{I_t}^{-}) < \min (\mu_{I_t}^+, \gamma_{I_t}^{+})$}
    \STATE $(\mu_{I_t}^-, \mu_{I_t}^+) \leftarrow (\max(\mu_{I_t}^-, \gamma_{I_t}^{-}) , \min (\mu_{I_t}^+, \gamma_{I_t}^{+}))$
    \ELSE
    \STATE $\left( \mu_{I_t}^-, \mu_{I_t}^+\right) \leftarrow \left( \frac{ \mu_{I_t}^- +  \mu_{I_t}^+ }{2} , \frac{\mu_{I_t}^- + \mu_{I_t}^+}{2} \right)$
    \ENDIF
    \ENDIF
    \ELSE 
   \STATE $I_t \leftarrow argmax_i\hspace{0.2em} b_i\cdot (\hat{\theta}_{i}^{T} x_t)$, \hspace{0.2em} $ \ni I_t \in S_{act}$
   \STATE Observe click as $r_{I_t} \in \{0, 1\}$
    \ENDIF
    \FORALL{agent $i \in S_{act}$}
    \IF{$\mu_i^+ < \max_{k\in S_{act}}\mu_{k}^-$}
    \STATE Remove $i$ from $S_{act}$
    \ENDIF
    \ENDFOR
\ENDFOR
\end{algorithmic}\end{small}
\end{algorithm}
\emph{ELinUCB-S} (Algorithm \ref{algo:linucb_base}) for single-slot allocation maintains a set of active agents $S_{act}$. At each round, algorithm evaluates whether an agent should be retained in $S_{act}$ or not. Once an agent is evicted from $S_{act}$, it can not be added back. For better understanding about the working of the algorithm, we virtually divide the \emph{LinUCB-S} into 4 subroutines: \romannumeral 1) Initialization (lines[1-7]) \romannumeral 2) Exploration (lines[11-20]) \romannumeral 3) Exploitation (lines[22-23]) \romannumeral 4) Elimination
(lines[24-26]).\footnote{Note that this is virtual division and proposed algorithms is not actually exploration-separated where initial rounds are only exploration as well as free and then exploitation where no update happens.} For each agent $i$, the algorithm maintains lower confidence bound (LCB) and upper confidence bound (UCB) as $\mu_i^-$ and $\mu_i^+$ respectively. \par
At each round $t$, the algorithm observes context $x_t$. It determines the index of agent $I_{t^{'}}$ whose turn is to display the ad based on round robin order, as stated in line[9]. The algorithm then checks if $I_{t^{'}} \in S_{act}$. If it evaluates to true the algorithm runs Exploration subroutine else Exploitation. In Exploration subroutine the algorithm allocates the slot to $I_{t^{'}}$, observes click $r_{I_t^{'}}$ and updates its parameters. The confidence bounds are updated if and only if the size of confidence interval decreases (line[18]). In Exploitation subroutine, the agent with the maximum estimated expected valuation among the agents in $S_{act}$ is allocated the slot and observes click $r_{I_t}$. It is important to note that no parameter is updated during Exploitation subroutine which is crucial for the ex-post monotonicity property. At the end of each round, Elimination subroutine is executed which removes the agents $j \in S_{act}$ from $S_{act}$ if UCB of agent $j$ is less than LCB of any other agent in $S_{act}$. 

The intuition driving the algorithm is after sufficient exploration the confidence interval becomes sufficiently small, hence the agents which are close to optimal continues to remain in $S_{act}$ and sub-optimal agents are eliminated.
\begin{algorithm}[!ht] \begin{small}
\caption{\emph{ELinUCB-SB}: LinUCB-based batch allocation rule for single-slot SSA}
\label{algo:linucb_SSA}
\algsetup{indent=0.75em}
\begin{algorithmic}[1]
\STATE \textbf{Inputs:} $n, T$, $\alpha \in \mathbbm{R}_{+}$, 
bid vector $b$, batch size $bs$
\STATE \textbf{Initialization:}  $S_{act} =$ $\mathcal{N}$,  $ x^{'} \leftarrow 0_{d\times 1}, T^{'} = \lfloor \frac{T}{bs}\rfloor$ 
\FORALL  {$i \in \mathcal{N}$}
    \STATE $A_i \leftarrow I_d$ (d-dimensional identity matrix) 
    \STATE $c_i \leftarrow 0_{d\times 1}$ (d-dimensional zero vector) 
    \STATE $\mu_i^+ \leftarrow b_i; \mu_i^- \leftarrow 0$ 
\ENDFOR

\FOR {$t^{'} = 1,2,3,\ldots, T^{'} $}
    \STATE $I_{t^{'}} \leftarrow 1+ (t^{'} -1) \mod n $ 
    \IF{$I_{t^{'}} \in S_{act}$}
    \FOR{$t = (t^{'}-1)bs, \ldots, (t^{'}\cdot bs - 1) $}
    \STATE  Observe context as $ x_{t} $
    \STATE $I_t \leftarrow I_{t^{'}}$, 
    \STATE $x^{'} \leftarrow((t-1)x^{'} + x_{t})/t$ (averaging over contexts)
    \STATE Observe click as $r_{I_t} \in \{0, 1\}$
    \STATE $A_{I_t} \leftarrow A_{I_t} + x_{t}x_{t}^\intercal$, $c_{I_t} \leftarrow c_{I_t} + r_{I_t}x_{{t}}$, $\hat{\theta}_{I_t} \leftarrow A_{I_t}^{-1}c_{I_t}$
    \ENDFOR
    \IF {$\mu_{I_t}^{-} < \mu_{I_t}^{+}$}
    \STATE $(\gamma_{I_t}^{-} , \gamma_{I_t}^{+}) \leftarrow b_{I_t}(\hat{\theta}_{I_t}^\intercal x^{'} \mp \alpha \sqrt{(x^{'})^\intercal A_{I_t}^{-1}x^{'}})$
    \IF {$\max(\mu_{I_t}^-, \gamma_{I_t}^{-}) < \min (\mu_{I_t}^+, \gamma_{I_t}^{+})$}
    \STATE $(\mu_{I_t}^-, \mu_{I_t}^+) \leftarrow (\max(\mu_{I_t}^-, \gamma_{I_t}^{-}) , \min (\mu_{I_t}^+, \gamma_{I_t}^{+}))$
    \ELSE
    \STATE $\left( \mu_{I_t}^-, \mu_{I_t}^+\right) \leftarrow \left( \frac{ \mu_{I_t}^- +  \mu_{I_t}^+ }{2} , \frac{\mu_{I_t}^- + \mu_{I_t}^+}{2} \right)$
    \ENDIF
    \ENDIF
   
    \ELSE
    \FOR{$t = (t^{'}-1)bs, \ldots, (t^{'}\cdot bs - 1) $}
    \STATE Observe $ x_{t} $
    \STATE $I_t \leftarrow argmax_i\hspace{0.2em} b_i \cdot (\hat{\theta}_{i}^{T} x_t)$, \hspace{0.2em} $ \ni I_t \in S_{act}$
    \STATE Observe click as $r_{I_t} \in \{0, 1\}$
    \ENDFOR
    \ENDIF
    
    \FORALL{agent $i \in S_{act}$}
    \IF{$\mu_i^+ < \max_{k\in S_{act}}\mu_{k}^-$}
    \STATE Remove $i$ from $S_{act}$
    \ENDIF
    \ENDFOR
   
\ENDFOR
\end{algorithmic} \end{small}
\end{algorithm}
\par
\subsection{Regret Analysis of \emph{ELinUCB-SB}}
Although the algorithm \emph{ELinUCB-S} seems promising, the dynamic and varying nature of contexts and its arrival order may lead to the elimination of an optimal agent. Hence, it may continue to allocate sub-optimal agents in subsequent rounds leading to high regret on specific instances, which is evident from our simulation of the algorithm (Fig.\ref{fig:sub3}). The updates in $\mu_i^+$, $\mu_i^-$ depend upon the context in such a way that $\mu_i^+$ is non-increasing and $\mu_i^-$ is non-decreasing, as stated and proved in Claim \ref{cl_2}. These updates being irreversible needs to be carefully handled to optimize regret. To counter this problem, we design \emph{ELinUCB-SB} (Algorithm \ref{algo:linucb_SSA}) in which we have introduced a subtle, yet important use of mini-batch. The algorithm \textit{ELinUCB-SB} allocates an agent for $bs$ number of rounds instead of one round. It follows similar rules for allocating agents and maintaining the active set $S_{act}$. It updates the bounds of agents by taking the average over the contexts arrived in $bs$ rounds. Updating the bounds over the average of context after the completion of batch allocation handles the variance in contexts and its arrivals, thus reducing the regret significantly. 

It can be shown that eventually, \emph{ELinUCB-SB} will eliminate all but one arm. The remaining arm will be the dominant arm in most of the contexts.  However, we can construct examples where this arm is not the best for at least one context, which has non-zero probability and thus leading to $O(T)$ regret. However, the round number at which it happens is generally very high, which we validate experimentally. Even though \emph{ELinUCB-SB} incurs linear regret theoretically, it performs well in experiments and has interesting monotonicity properties; the proofs we leverage while designing ex-post truthful ConMAB mechanism with sub-linear regret in the next section.

\subsection{Monotonicity of \emph{ELinUCB-S}}
We now prove ex-post monotonicity property for the proposed allocation rule.

For a fixed sequence of context-arrivals $\{x_t\}_t$, and click realization $\rho$, let $S_{act}(b,t)$ be the set of active agents in the beginning of round $t$ when agents  bid  $b=(b_i,b_{-i})$. For each agent $i$, let $\mu_i^-(b, t)$ and $\mu_i^+(b, t)$ be the values of $\mu_i^-$ and $\mu_i^+$ in the round $t$ and similarly when agents bid $b'$. We prove ex-post monotonicity with the following claims.
\begin{claim}
\label{cl_1}
For fixed context-arrivals $\{x_t\}_t$ and click realization $\rho$, let two bid vectors be \emph{b} and $b^{\prime}$, $\text{ then}$ for each round $t$, if  $i  \in S_{act}(b, t) \cap S_{act}( b^{'}, t)$, then:
    \begin{equation*}
         \mu_{ i }^- (b,t ) / b_{ i } = \mu_{ i }^- ( b^{ \prime }, t ) / b_{ i }^{ \prime }  \text{ and }
         \mu_{ i }^+ ( b, t ) / b_{ i } = \mu_{ i }^+ (  b^{ \prime }, t) / b_{ i }^{ \prime } 
    \end{equation*}
 \end{claim}
\begin{proof}
$\mu_i^+$ and $\mu_i^-$ are updated only in Exploration subroutine which is based on round-robin order and hence does not depend on bid. Thus, the claim follows from the fact that contexts and click realizations are fixed.
\end{proof}

\begin{claim}
\label{cl_2}
For a fixed bid vector $b$, and each agent $i: \mu_i^- \leq \mu_i^+$, then for all $(t-1, t)$ consecutive rounds $\mu_i^-$ is non-decreasing and $\mu_i^+$ is non-increasing. 
\end{claim}
\begin{proof}
From lines[15-20] of the Algorithm \ref{algo:linucb_base}, we have: $\mu_{ i }^- (  b, t-1 ) \leq \mu_{ i }^- (  b,t ) \leq \mu_{ i }^+ (  b,t ) \leq \mu_{ i }^+ ( b, t - 1  )$. Hence the claim holds.
\end{proof}

\begin{claim}
\label{cl_3}
For any two bid vectors $b^+$ and $b$, where $b_i^+ \geq b_i$, $b^+_j = b_j$ $\forall j \neq i$  and $i \in S_{act}(b^+, \tau) \cap S_{act}(b, \tau)$, then $\forall \tau \in \{1,2,\ldots, T\}$, $S_{act}(b^+, \tau) \subseteq S_{act}(b, \tau)$ holds.
\end{claim}
\begin{proof}
The condition $i \in S_{act}(b^+, \tau) \cap S_{act}(b, \tau)$ implies that if $i \in S_{act}(b^+, \tau)$, then $i \in S_{act}(b, \tau)$, hence satisfying the claim for $i$. For $j \neq i$, we will use induction on $t$. The claim trivially holds for $t= 1$. Let, $t \le \tau$ be the last round such that $S_{act}(b^+, t) = S_{act}(b, t)$ and $S_{act}(b^+, t+1) \neq S_{act}(b, t+1)$. In this case, we prove that: $\forall j\ne i, j \in S_{act}(b^+, t+1) \implies j \in S_{act}(b, t+1)$. \\
Since, $j \ne i$, $\mu_j^+(b^+, t+1) = \mu_j^+(b,t+1)$, $\mu_z^-(b^+, t+1) = \mu_z^-(b,t+1)\ \forall z \ne i$, and $\mu_i^-(b^+, t+1) \ge \mu_i^-(b,t+1)$ from Claim \ref{cl_1}. Thus,
\begin{align*}
    \mu_j^+(b^+,t+1) &> \max_{z \in S_{act}(b^+, t)} \mu_z^-(b^+,t+1)\\
    &\geq \max_{z \in S_{act}(b^+, t)} \mu_z^-(b,t+1)\\
    \implies \mu_j^+(b,t+1) &> \max_{z \in S_{act}(b, t)} \mu_z^-(b,t+1)
\end{align*}
(Since $S_{act}(b, t) = S_{act}(b^+,t)$). Hence, $j \in S_{act}(b, t+1)$.
From, induction hypothesis, $\forall t'$ s.t. $\tau > t' \ge t$, $S_{act}(b^+, t') \subseteq S_{act}(b, t')$. We will now prove that $S_{act}(b^+, t'+1) \subseteq S_{act}(b, t'+1)$.

Consider any $j \in S_{act}(b^+, t') \cap S_{act}(b, t')$: we will prove that if $j \in S_{act}(b^+, t'+1)$, then $j \in S_{act}(b, t'+1)$.

Since $j \in S_{act}(b^+, t') \cap S_{act}(b, t')$, $\mu_j^+(b^+, t'+1) = \mu_j^+(b, t'+1)$. Also, $\forall z \in S_{act}(b, t') \cap S_{act}(b^+, t')$ and $z \ne i$, $\mu_z^-(b^+, t'+1) = \mu_z^-(b,t'+1)$. Further, $\forall z$ such that $z \in S_{act}(b, t')$ but $z \notin S_{act}(b^+, t')$, $\exists l \in S_{act}(b^+, t')$ such that $\mu_z^+(b^+, t'+1) < \mu_l^-(b^+, t'+1) \implies \mu_z^-(b^+, t'+1) < \mu_l^-(b^+, t'+1)$. Thus, $\max_{z \in S_{act}(b, t'+1)}\mu_z^-(b^+, t'+1) \le \max_{z \in S_{act}(b^+, t'+1)}\mu_z^-(b^+, t'+1)$. 
Thus, $j \in S_{act}(b^+, t' + 1)$ implies
\begin{align*}
 \mu_j^+(b^+. t'+1) &\ge \max_{z \in S_{act}(b^+, t')} \mu_z^-(b^+, t'+1)\\
\implies \mu_j^+(b, t'+1) &\ge \max_{z \in S_{act}(b, t')} \mu_z^-(b^+, t'+1)\\
\implies \mu_j^+(b, t'+1) &\ge \max_{z \in S_{act}(b, t')} \mu_z^-(b, t'+1)
\end{align*}
\end{proof}

\begin{claim}
\label{cl_4}
For fixed context-arrivals, fixed click realizations, and fixed bids of the agents except $i$, that is, for a fixed $b_{-i}$, if $i \in S_{act}(t,b)$, then $i \in S_{act}(t,b^+)$ where $b=(b_i,b_{-i})$ and $b^+=(b_i^+,b_{-i})$; $b_i^+>b_i$.
\end{claim}
\begin{proof}
Let $\tau^*\ge 1$ be the last round for $i$ s.t. it is in active set with both bids. From Claim 1, $\mu_i^+(b^+, \tau^*) = \frac{b_i^+}{b_i}\mu_i^+(b, \tau^*)> \mu_i^+(b, \tau^*)$
as $i\in S_{act}(b, \tau^*) \cap S_{act}(b^+, \tau^*) $.
As the context-arrivals, click realizations and bids of the remaining agents are fixed, 
if agent $i$ becomes inactive with bid $b_i^+$ then
\begin{align*}
    \mu_i^+(b^+, \tau^*) &<  \max_{k \in S_{act}(b^+, \tau^*-1)} \mu_k^-(b^+, \tau^*)\\
\implies \mu_i^+(b, \tau^*) &<  \max_{k \in S_{act}(b^+, \tau^*-1)} \mu_k^-(b^+, \tau^*)\\
\implies \mu_i^+(b, \tau^*) &<  \max_{k \in S_{act}(b, \tau^*-1)} \mu_k^-(b, \tau^*).
\end{align*}
The last line follows from Claim \ref{cl_3}. Thus, $i$ is also inactive with bid $b_i$. 
\end{proof}

\begin{theorem}
\label{thm:linucb_bsln_mon}
The allocation rule induced by ELinUCB-S (Algorithm \ref{algo:linucb_base}) is ex-post monotone.

\end{theorem}
\begin{proof}
For a fixed context-arrivals $\{x_t\}_t$, click realization $\rho$, bids of agents except $i$, i.e., $b_{-i}$ and two possible bids $b_i^+ > b_i$, let $\tau$ and $\tau^+$ be the last round for $i$ s.t. $i$ is in active set with bids $b_i$ and $b_i^+$ respectively. From Claim \ref{cl_4}, $\tau^+ \geq \tau$. Thus, $i$ will receive more number of rounds with bid $b_i^+$ as compared with bid $b_i$.
\end{proof}

\begin{Proposition}
\label{prp:linucb_s_mon}
The allocation rule induced by ELinUCB-SB (Algorithm \ref{algo:linucb_SSA}) is ex-post monotone.
\end{Proposition}
\begin{proof}
The difference between Algorithm \ref{algo:linucb_base} and Algorithm \ref{algo:linucb_SSA} is the introduction of batch allocation. In Algorithm \ref{algo:linucb_SSA} the unit of one round is equivalent to $bs$ rounds in Algorithm \ref{algo:linucb_base}. Hence, by replacing variable $t$ with $t^{'}$ where $t \in \{1,2, \ldots, T\}$ and $t^{'} \in \{1,2, \ldots, \lfloor \frac{T}{bs}\rfloor\}$ will satisfy all the claims (Claim 1-4). Thus, \emph{ELinUCB-SB} (Algorithm \ref{algo:linucb_SSA}) is still ex-post monotone. 
\end{proof}

\subsection*{\emph{M-ELinUCB-SB}}
We now propose the following mechanism \emph{M-ELinUCB-SB} for the single-slot SSA. A mechanism is defined as $\mathcal{M}=(\mathcal{A}, \mathcal{P})$. The outline of both mechanisms is defined in Mechanism \ref{mech:mech1}. For both the mechanisms, we apply the resampling procedure \cite{Bab15} on the bids and the allocation in both cases are based on the modified bids, where $\delta$ is resampling parameter. For \emph{M-ELinUCB-SB}, $\mathcal{A}$ is given by \emph{ELinUCB-SB}. The payment $\mathcal{P}$ at round $t$, corresponding context $x_t$, $\forall i \in \mathcal{N}$ is given by $ p_{i,t} = b _ { i } \cdot (\mathbbm{1}\{I_t = i\}) \text{ if } \eta _ { i } = 1$ and $ p_{i,t} = b _ { i } \cdot (\mathbbm{1}\{I_t = i\}) \cdot  (1 - \frac { 1 } { \delta }) \text { if } \eta _ { i } < 1$.     
\begin{algorithm} \begin{small}
    \caption{\emph{M-ELinUCB-SB}: LinUCB-based ex-post truthful mechanism }
    \label{mech:mech1}
        \begin{algorithmic}[1]
            \STATE \textbf{Input:} bid vector $b$, resampling parameter $\delta$
            \STATE Run self-resampling procedure on bid vector $b$, obtain modified bid vector $y =(y_1,\ldots, y_n)$, $\eta = (\eta_1, \ldots, \eta_n$)  
            \STATE Allocate according to $\mathcal{A}(y, t)$
            \STATE For each agent $i$, assign payment $ p_{i,t} = b _ { i } \cdot \mathcal { A } _ { i,t } ( y, t) \cdot \left\{ \begin{array} { l l } { 1 } & { \text { if } \eta _ { i } = 1 } \\ { 1 - \frac { 1 } { \delta } } & { \text { if } \eta _ { i } < 1 } \end{array} \right.$
        \end{algorithmic} \end{small}
    \end{algorithm}


\subsection{\emph{M-ELinUCB-SB}: Game Theoretic Analysis}
\begin{theorem}
\label{thm:M-LinUCB-SB_EPIC}
  \emph{M-ELinUCB-SB} is ex-post incentive compatible (EPIC) and ex-post individually rational (EPIR) mechanism.
 \end{theorem}
\begin{proof}
  The result follows from Theorem \ref{thm:babaioff} and by ex-post monotonicity of $\mathcal{A}$ defined in Algorithm \ref{algo:linucb_SSA}.
 \end{proof}

\section{\emph{M-SupLinUCB-S}: Truthful ConMAB Mechanism 2}
As the mechanism with allocation rule in Algorithm \ref{algo:linucb_SSA} can incur linear regret, in this section, we propose a new ConMAB mechanism for SSA that achieves sub-linear regret. First, we explain how we adapt SupLinUCB \cite{Chu11} for SSA to derive an ex-post monotone allocation algorithm SupLinUCB-S in the next subsection. In Section \ref{ssec:suplinucb_reg}, we prove the regret bound on SupLinUCB-S. Then we prove the monotonicity of it. Finally, we design a truthful mechanism M-SupLinUCB-S. 
\subsection{\emph{SupLinUCB-S}}
Chu et al. \cite{Chu11} proposed SupLinUCB for contextual MAB settings with linear payoffs. 
First let us emphasize the major differences between SupLinUCB (\cite{Chu11}) and SupLinUCB-S (proposed here, Algorithm \ref{algo:SupLinUCB-S}). (i) Chu et al. have considered a common $\theta$ to be learned across all the agent and for each round $t$ the contexts are different for each agent whereas in our setting we have independent $\theta_i$ to be learned for each agent $i$ whereas the context across each agent is same. (ii) We have adapted their algorithm for auction setting such that it satisfies \textit{ex-post monotonicity} property, which is necessary to design ex-post truthful mechanisms. Our algorithm is presented in \ref{algo:SupLinUCB-S}.

\begin{algorithm}[!ht] 
\caption{\emph{SupLinUCB-S:(Adapted from SupLinUCB by \cite{Chu11} to satisfy monotonicity property)}}
\label{algo:SupLinUCB-S}
\algsetup{indent=0.75em}
\begin{small}
\begin{algorithmic}[1]
\STATE Initialization: $S \leftarrow \ln T$, $\Psi_{i,t}^s \leftarrow \phi \text{ for all } s \in [\ln T]$
\FOR{t = 1,2,\ldots, T} 
\STATE $s \leftarrow 1 \text{ and } \hat{A}_1 \leftarrow \mathcal{N} $
\STATE $j \leftarrow 1 + (t \text{ mod } n)$
\REPEAT
\STATE Use \textit{BaseLinUCB-S} with $\{\Psi_{i,t}^s\}_{i\in \mathcal{N}}$ and context vector $x_t$ to calculate the width $w_{i,t}^s$ and upper confidence bound $ucb_{i,t}^s$ $=  (\hat{r}_{i,t}^s+ w_{i,t}^s)$, $\forall i \in \hat{A}_s$
\IF {$j \in \hat{A}_s \text{ and }  w_{j,t}^s > 2^{-s}$}
\STATE Select $I_t = j$
\STATE Update the index sets at all levels:
\\
$\Psi_{i,t+1}^{s^{\prime}} \leftarrow 
\left\{ \begin{array} { l l } { \Psi_{i,t}^{s^{\prime}} \cup \{t\} } & { \text { if } s=s^{\prime} } 
\\ 
{\Psi_{i,t}^{s^{\prime}} } & { \text {otherwise} } \end{array} \right.$
\ELSIF{$w_{i,t}^s \leq \frac{1}{\sqrt{T}}, \forall i \in \hat{A}_s$}
\STATE Select $I_t = argmax_{i \in \hat{A}_s} b_i \cdot(\hat{r}_{i,t}^s+ w_{i,t}^s)$
\STATE Update index sets at all levels for $I_t$:
\\
$\Psi_{I_t,t+1}^{s^{\prime}}\leftarrow \Psi_{I_t,t}^{s^{\prime}}$, $\forall s^{\prime} \in [S]$
\ELSIF{$w_{i,t}^s \leq 2^{-s}, \forall i \in \hat{A}_s$}
\STATE $\hat{A}_{s+1} \leftarrow \{ i \in \hat{A}_s | b_i \cdot( \hat{r}_{i,t}^s+ w_{i,t}^s) \geq \max_{a \in \hat{A}_s}b_a \cdot(\hat{r}_{a,t}^s+ w_{a,t}^s) - 2^{1-s} \}$
\STATE $s \leftarrow s+1$
\ELSE
\STATE Select  $I_t = argmax_{i \in \hat{A}_s} b_i \cdot (\hat{r}_{i,t}^s+ w_{i,t}^s)$
\ENDIF
\UNTIL{$I_t$ is selected}
\ENDFOR
\end{algorithmic}\end{small}
\end{algorithm}

\begin{algorithm}[!ht] 
\caption{\emph{BaseLinUCB-S: (Adapted from BaseLinUCB by \cite{Chu11})}}
\label{algo:BaseLinUCB-S}
\algsetup{indent=0.75em}
\begin{small}
\begin{algorithmic}[1]
\STATE \textbf{Inputs:} $\alpha \in \mathbbm{R}_{+}$, $\Psi_{i,t} \subseteq \{1,2,\ldots, t-1\}$
\STATE  $A_{i,t} \leftarrow I_d + \sum_{\tau \in \Psi_{i,t}} x_{\tau}^\intercal x_{\tau}$
\STATE $c_{i,t} \leftarrow \sum_{\tau \in \Psi_{i,t}} r_{i,\tau} x_{\tau}$
\STATE $\theta_{i,t} \leftarrow A_{i,t}^{-1}c_{i,t}$
\STATE Observe context vector as $x_{t} \in [0,1]^d$
\FOR {$i \in \mathcal{N}$}
\STATE $w_{i,t}^s \leftarrow \alpha \sqrt{x_{t}^\intercal A_{i,t}^{-1}x_{t}}$
\STATE $\hat{r}_{i,t}^s \leftarrow \theta_{i,t}^{\intercal}x_{t}$
\ENDFOR
\end{algorithmic}\end{small}
\end{algorithm}

In the next section, we prove the regret bounds on \emph{SupLinUCB-S}, highlighting the steps which differ from regret analysis of \emph{SupLinUCB}. 
\subsection{Regret Analysis of \emph{SupLinUCB-S}}
\label{ssec:suplinucb_reg}

For convenience, let $s_{i,t} = \sqrt{x_{t}^\intercal A_{i,t}^{-1}x_{t}}$, $ucb_{i,t}^s =  (\hat{r}_{i,t}^s+ w_{i,t}^s)$ and $0 < b_i <=1, \forall i \in \mathcal{N}$. For all round $t$, stage $s$ and given context $x_t$,  $i^*_t(x_t) = argmax_{i \in \hat{A}_s} b_i \cdot \mathbbm{E}[r_{i,t}|x_t]$. The regret analysis is along the similar lines with \cite{Chu11} with changes deemed necessary to incorporate in our setting. In Lemmas \ref{lemma_1},  \ref{lemma_2}, and  \ref{lemma_3}, we need to work for each agent as we have different $\theta_i$s for different agents. 
\begin{lemma}
\label{lemma_1}
(Lemma 2, \cite{Chu11}) For each $s \in [S]$ and $i \in \mathcal{N}$, suppose $\psi_{i,t+1}^s = \psi_{i,t}^s \cup \{t\}$. Then, eigenvalues of $A_{i,t}$ can be arranged so that $\lambda_{i,t}^j \leq \lambda_{i, t+1}^j$, for all $j$ and 
\begin{equation*}
    s_{i,t}^2 \leq 10 \sum_{j=1}^d \frac{\lambda_{i,t+1}^j - \lambda_{i,t}^j}{\lambda_{i,t}^j}
\end{equation*}
\end{lemma}
\begin{lemma}
\label{lemma_2}
(Lemma 3, \cite{Chu11}) Using notation in \emph{BaseLinUCB-S} and assuming $|\psi_{i,T+1}^s| \geq 2$, we have 
\begin{equation*}
    \sum_{t \in \psi_{i,T+1}^s} s_{i,t} \leq 5\sqrt{d|\psi_{i,T+1}^s|\ln{|\psi_{i,T+1}^s|}}
\end{equation*}
\end{lemma}
\begin{lemma}
\label{lemma_3}
(Lemma 4, \cite{Chu11}). For each $s \in [S]$, each $t \in [T]$, and any fixed sequence of feature vectors $x_t$, with $t \in \psi_{I_t,t}^s$, the corresponding rewards $r_{I_t,t}$ are independent random variables such that $\mathbbm{E}[r_{I_t,t}] = \theta_i^{\intercal} x_{t}$.
\end{lemma}
In our settings, rewards of the arms also have bid component which plays an important role, thus we need the following lemma.
\begin{lemma}
\label{lemma_4}
With probability $1-\kappa S$, for any $t \in [T]$ and any $s \in [S]$, the following hold:
\begin{enumerate}
    \item $b_i\cdot ucb_{i,t}^s - 2\cdot w_{i,t}^s \leq b_i \cdot \mathbbm{E}[r_{i,t}] \leq b_i \cdot ucb_{i,t}^s$ for all $i$
    \item $i^*_t(x_t) \in \hat{A}_s$
    \item $b_{i^*_t(x_t)}\cdot \mathbbm{E}[r_{i^*(x_t),t}] - b_{i}\cdot \mathbbm{E}[r_{i,t}] \leq 2^{3-s}$
\end{enumerate}
\end{lemma}
\begin{proof}
From  Lemma 15 of \cite{Auer03}, we have, $ ucb_{i,t}^s - 2\cdot w_{i,t}^s \leq \mathbbm{E}[r_{i,t}] \leq ucb_{i,t}^s$ for all $i$. As $b_i > 0$, after multiplying with $b_i$ inequality still holds, i.e., $b_i\cdot ucb_{i,t}^s - 2\cdot b_i w_{i,t}^s \leq b_i \cdot \mathbbm{E}[r_{i,t}] \leq b_i \cdot ucb_{i,t}^s$. From our assumption $b_i \leq 1$, hence the first part holds.
\\ \\
The lemma trivially holds for $s=1$. For $s >1$, $\hat{A}_s \subseteq \hat{A}_{s-1} $ and from the algorithm it is clear that $w_{i,t}^s \leq 2^{-(s-1)}$ and $w_{i^*_t(x_t)}^s \leq 2^{-(s-1)}$. From part 1 of the lemma and using the above fact, for any $j \in \hat{A}_s$ we have $b_{i^*_t(x_t)} ucb_{i^*_t(x_t), t}^{(s-1)} \geq b_{i^*_t(x_t)}\mathbbm{E}[r_{i^*_t(x_t), t}]$ and $b_j \mathbbm{E}[r_{j,t}] \geq b_j ucb_{j,t}^s - 2\cdot 2^{-(s-1)}$. From definition, $b_{i^*_t(x_t)} \mathbbm{E}[r_{i^*_t(x_t), t}] \geq b_j \mathbbm{E}[r_{j,t}]$. Using this and above inequalities, agent $i^*_t(x_t)$ will belong to $\hat{A}_s, \forall s$ (i.e., will never be eliminated for context $x_t$) due to the rule defined in Line[14], Algorithm \ref{algo:SupLinUCB-S}. Hence part 2 of the lemma is proved. 
\\ \\
From Line[14] Algorithm \ref{algo:SupLinUCB-S}, $b_i ucb_{i,t}^s \geq b_{i^*_t(x_t)} ucb_{i^*_t(x_t), t}^s - 2\cdot 2^{-(s-1)}$. Using part 1 of the lemma and above inequality the proof of part 3 follows.
\end{proof}
\begin{lemma}
\label{lemma_5}
(Lemma 6, \cite{Chu11}) For all $s \in [S]$ and $i \in \mathcal{N}$,
\begin{equation*}
    |\psi_{i, T+1}^s| \leq 5\cdot 2^s(1+\alpha^2) \sqrt{d|\psi_{i, t+1}^s|}
\end{equation*}
\end{lemma}

All the above can be summarized as the following theorem. 
\begin{theorem}
\label{thm:suplinucb_reg}
SupLinUCB-S has regret $O(n^2\sqrt{dT\ln{T})}$ with probability at least $1-\kappa$ if it is run with $\alpha = \sqrt{\frac{1}{2}\ln{\frac{2nT}{\kappa}}}$.
\end{theorem}
\begin{proof}
The proof is similar to the proof of Theorem 6 of \cite{Auer03} but requires additional terms to consolidate the difference between the algorithms, problem setting, and regret definition. We have restricted the learning during round-robin ordering only Lines[7-9] due to which we have an additional decision rule for agent selection as in Line[17]. Note that this additional rule was not in \cite{Chu11}. Hence the main challenge is to bound the number of rounds agent selection, which is done using this decision rule. (We refer it $\psi_{ext}^s$ in our analysis.)

Let $\psi_0$ be the set of rounds in which the agent was selected in Lines[10-12]. Let $\psi_{ext}^s$ be the set of rounds agent was selected in Lines[16-17] and $\psi_{T+1}^s = \bigcup_i \psi_{i, T+1}^s$.

\begin{claim}
At each stage $s$, $|\psi_{est}^s| <= (n-1)\cdot |\psi_{T+1}^s|$.
\end{claim}
For any stage $s$, let us take case of $n$ consecutive rounds. Let us assume for each of the $n$ rounds, selection of agent is done in Lines[16-17]. But note that selection of agent in this decision block is done if and only if there exist an agent $k$ such that $j\neq k$ (where $j$ is designated agent for the round) and $w_{k,\cdot}^s > 2^{-s}$. But after $n$ consecutive rounds each agent has got its designated round once (Line[4]). Hence, if for some agent $k$, $w_{k,\cdot}^s > 2^{-s}$, then this agent $k$ should be selected on its designated round. Hence our assumption of selection of agent in Lines[16-17] for $n$ consecutive round is wrong. From above reasoning, it is clear that at least for one round out of $n$ rounds, one of the agent must be selected at its designated round. Hence, at most for $n-1$ rounds agent is selected in Line[16-17] out of $n$ rounds, until condition in Line[10] is achieved. Thus, we can say that at each stage $s$, $|\psi_{est}^s| <= (n-1)\cdot |\psi_{T+1}^s|$. 
\\
 As $2^{-S} \leq 1/ \sqrt{T}$, we have $\{1,\ldots,T\} = \psi_0 \cup \bigcup_s \psi_{T+1}^s \bigcup_s \cup \psi_{est}^s$. Using the claim and the lemmas, 
\begin{align*}
    \mathbbm{R}_T = &  \sum_{t=1}^T [b_{i^*_t(x_t)}\mathbbm{E}[r_{i^*_t(x_t),t}]- b_{I_t}\mathbbm{E}[r_{I_t,t}]]\\
    = & \sum_{t\in\psi_0} [b_{i^*_t(x_t)}\mathbbm{E}[r_{i^*_t(x_t),t}]- b_{I_t}\mathbbm{E}[r_{I_t,t}]] \\
    & + \sum_{s=1}^S \bigg[\sum_{t \in \psi_{T+1}^s}[b_{i^*_t(x_t)}\mathbbm{E}[r_{i^*_t(x_t),t}]- b_{I_t}\mathbbm{E}[r_{I_t,t}]]\\
    & + \sum_{t\in \psi_{est}^s}[b_{i^*_t(x_t)}\mathbbm{E}[r_{i^*_t(x_t),t}]- b_{I_t}\mathbbm{E}[r_{I_t,t}]] \bigg]\\
    \leq & \frac{2}{\sqrt{T}}|\psi_0| + \sum_{s=1}^S n\cdot \sum_{t \in \psi_{T+1}^s}[b_{i^*_t(x_t)}\mathbbm{E}[r_{i^*_t(x_t),t}]-
    b_{I_t}\mathbbm{E}[r_{I_t,t}]]\\
    = & \frac{2}{\sqrt{T}}|\psi_0| + \sum_{s=1}^S n \sum_{i}^{|\mathcal{N}|}\sum_{t \in \psi_{i, T+1}^s}[b_{i^*_t(x_t)}\mathbbm{E}[r_{i^*_t(x_t),t}]- b_{I_t}\mathbbm{E}[r_{I_t,t}]]\\
    \leq & \frac{2}{\sqrt{T}}|\psi_0| + n \sum_{i}^{|\mathcal{N}|} \sum_{s=1}^S 8\cdot 2^{-s}\cdot |\psi_{i, T+1}^s|\\
    \leq & \frac{2}{\sqrt{T}}|\psi_0|+ n \sum_{i}^{|\mathcal{N}|} \sum_{s=1}^S 40\cdot(1+\ln{(2Tn/\kappa)})\cdot \sqrt{d|\psi_{i,T+1}^s|}\\
    \leq & \frac{2}{\sqrt{T}}|\psi_0|+ n \sum_{i}^{|\mathcal{N}|} 40\cdot(1+\ln{(2Tn/\kappa)}\cdot \sqrt{STd}\\
    \leq & 2\sqrt{T} + 40n^2\cdot(1+\ln{(2Tn/\kappa)}\cdot \sqrt{STd}
\end{align*}
\end{proof}

\begin{theorem}
The allocation rule induced by SupLinUCB-S (Algorithm \ref{algo:SupLinUCB-S}) is ex-post monotone.
\end{theorem}
\begin{proof}
The allocation rules Algorithm \ref{algo:linucb_base} and Algorithm \ref{algo:SupLinUCB-S} are similar in the way it learns and eliminates agents. Both algorithms learn only when a designated agent is selected based on round-robin ordering, and the elimination is based on bids, \emph{UCB} and \emph{LCB} estimates. The difference between elimination rules is the need for the width of an agent to reach threshold $2^{-s}$ at stage $s$. Due to the above similarities, the proof follows on the similar lines of the proof of Theorem \ref{thm:linucb_bsln_mon}, and hence we skip it for ease of exposition.
\end{proof}
\subsection*{\emph{M-SupLinUCB-S}}
The mechanism \emph{M-SupLinUCB-S} follows the same structure as that of mechanism \emph{M-ELinUCB-SB}. The only change is that the allocation rule $\mathcal{A}$ is given by \emph{SupLinUCB-S} (Algorithm \ref{algo:SupLinUCB-S}). 
\subsection{\emph{M-SupLinUCB-S}: Game-Theoretic Analysis}

\begin{theorem}
\label{thm:M-SupLinUCB_EPIC}
  \emph{M-SupLinUCB-S} is ex-post incentive compatible (EPIC) and ex-post individually rational (EPIR) mechanism.
 \end{theorem}
\begin{proof}
  The result follows from Theorem \ref{thm:babaioff} and by ex-post monotonicity of $\mathcal{A}$ defined in Algorithm \ref{algo:SupLinUCB-S}.
 \end{proof}

\section{Experimental Analysis}
\label{sec:empirical_analysis}
\subsection{Data Preparation}
Our simulated data follow the structure and information availability found in a real-world system.
Typically, a center has access to user features such as gender, age, geographic features, device model, and behavioral categories (which summarizes the user’s past preferences), which constitute the context. Note that each of the stated features can be discretized.  Considering the above facts, we created the corpus of users $\chi$: with $d=4$, for each feature, we randomly select $4$ possible different values from $0$ to $100$, and then by taking all possible combination of features, we generated random 256 ($4^4$) users. We normalize each $x \in \chi$ such that $x \in [0,1]^d$, with $||x||_2 = 1$ and store these normalized contexts as a database $\chi_{db}$. 

We then select $x_t$ uniformly at random from the context database $\chi_{db}$ for each round to generate a stochastic context. For each agent (advertiser), we generate $\theta_i \sim U([0,1]^d)$  and then normalize it s.t. $||\theta_i||_2 = 1$. To simulate the clicks, at round $t$ with the sampled $x_t$, we generate a click $r_{i,t}$ from Bernoulli distribution with parameter $\theta_i^Tx_t $.
We conduct  experiments for 40 iterations, and for each iteration, we randomly generate a sequence of contexts from $\chi_{db}$ for $T=10^6$ rounds. We generate valuation of agent $i$ for a click to be $v_i$ sampled from uniform distribution $[0,1]$ and assume the agents bid truthfully; due to truthfulness properties of our mechanisms.
\subsection{Results and Comparison}
\begin{figure}
\centering
\begin{subfigure}{.5\columnwidth}
\centering
\includegraphics[width=\columnwidth, height=4cm]{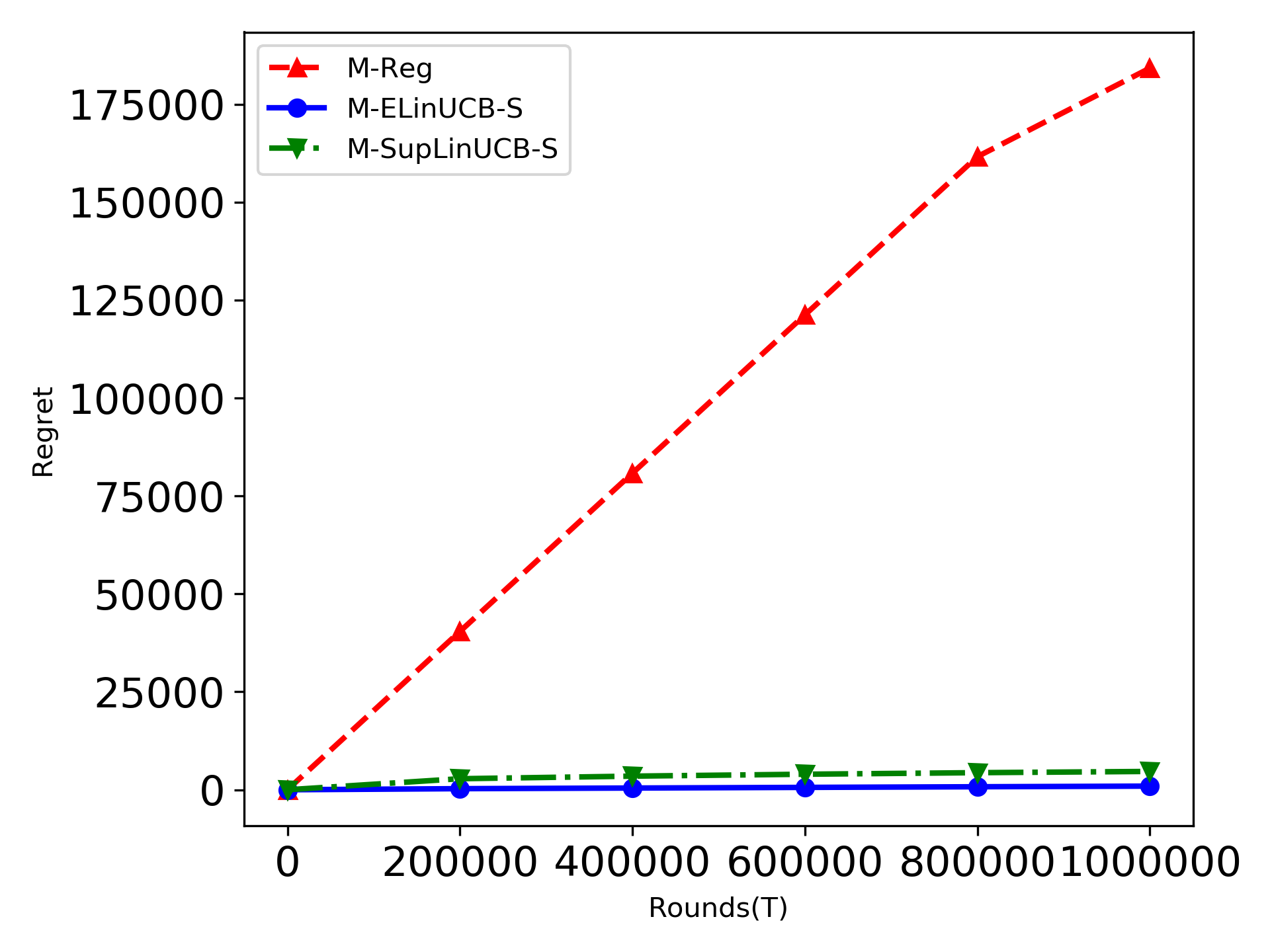}
\caption{Regret vs Rounds (T)}
\label{fig:sub1}
\end{subfigure}%
\begin{subfigure}{.5\columnwidth}
\centering
\includegraphics[width=\columnwidth, height=4cm]{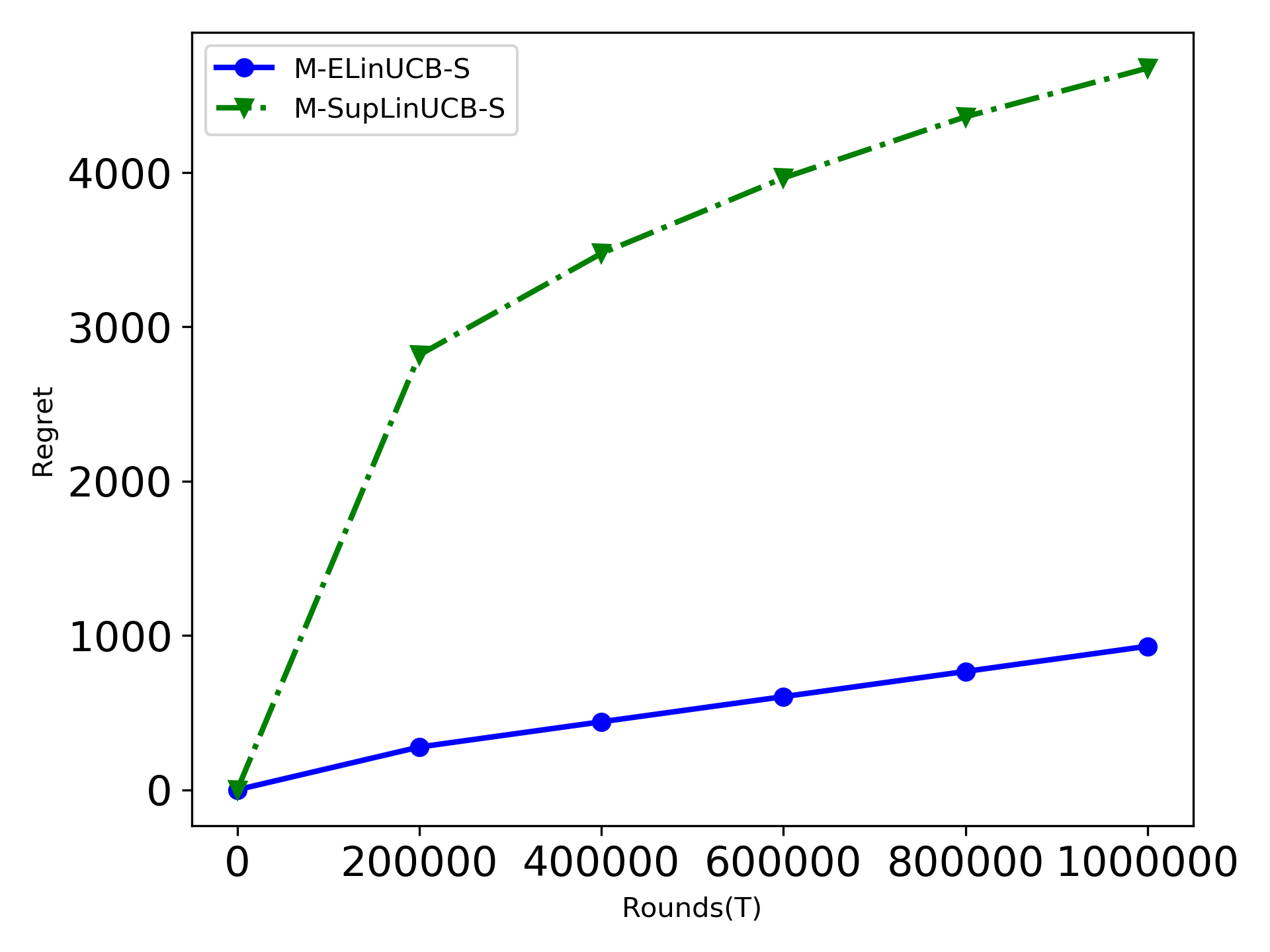}
\caption{Regret vs Rounds (T)}
\label{fig:sub2}
\end{subfigure}
\begin{subfigure}{\columnwidth}
\centering
\includegraphics[width=\columnwidth, height=6cm]{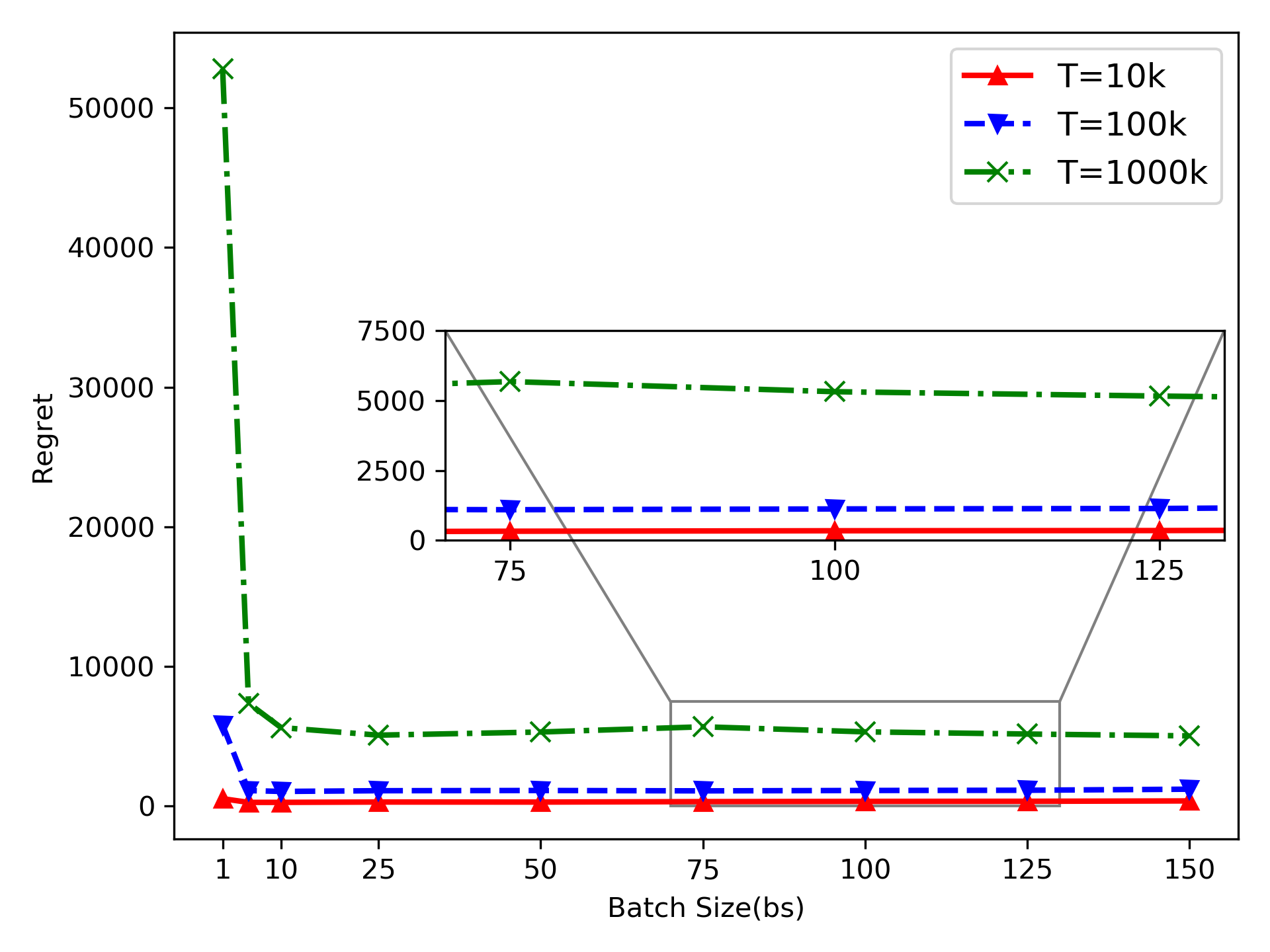}
\caption{Regret vs Batch Size (bs)}
\label{fig:sub3}
\end{subfigure}
\caption{Regret comparisons}
\label{fig:Comparison}
\end{figure}
From our experimentation, we found learning parameter $\alpha = 1$ and batch size $bs = 100$ to be suitable for \emph{M-ELinUCB-SB}. The metric of comparison between the mechanisms is regret, which is averaged over 40 iterations. Fig.\ref{fig:sub1} compares the regret of \emph{M-Reg}, \emph{M-ELinUCB-SB}, and \emph{M-SupLinUCB-SB} for $n=7$. When $n >7$, \emph{M-Reg} becomes infeasible as the number of exploration rounds $\lambda$ exceeds the total number of rounds $T$, for $T=10^6$. In terms of regret, it is evident that both the mechanisms \emph{M-ELinUCB-SB} and  \emph{M-SupLinUCB-SB} outperform \emph{M-Reg} by a very large margin. Fig.\ref{fig:sub2}  highlights difference in experimental regret \emph{M-ELinUCB-SB} and  \emph{M-SupLinUCB-SB} (it is zoomed version from Fig. \ref{fig:sub1}). We can see that \emph{M-ELinUCB-SB} experimentally performs approximately $5$ times better than \emph{M-SupLinUCB-SB}; albeit the results are validated only on the randomly generated 256 contexts ($\chi_{db}$). 

Though in theory, \emph{M-ELinUCB-SB} has the worst regret, from simulations, the slope being very small, for reasonable values of $T$ it outperforms \emph{M-Reg}. Our experiments  show that \emph{M-ELinUCB-SB} and  \emph{M-SupLinUCB-SB}, both have nearly negligible regret as compared to \emph{M-Reg}.  Fig.\ref{fig:sub3} compares the regret incurred by \emph{M-ELinUCB-SB} with varying batch size $bs \in \{1, 5, 10, 25, 50, 75, 100,$ $125, 150\}$ for $T \in \{10k, 100k, 1000k\}$. From the figure, it is easy to infer the significant improvement in regret when we move from batch size $bs=1$ to greater batch sizes. One may need to tune $bs$ based on different experimental setup.

\section{Conclusion}
\label{sec:con}
We believe that ours is the first attempt to design a non-exploration-separated ConMAB mechanism. We focused on designing ConMAB mechanisms for sponsored search auction (SSA). For a single-slot, we first designed LinUCB-based ex-post monotone allocation rule \emph{ELinUCB-S}. We show that the introduction of batch size in \emph{ELinUCB-S} significantly improves the regret while satisfying the ex-post monotone property. With this observation, we present \emph{ELinUCB-SB}. Through simulations, we see that in practice, it performs better for regret; however, theoretically, it may incur linear regret on carefully chosen contexts. To achieve sub-linear regret, we proposed another ex-post monotone allocation rule, \emph{SupLinUCB-S}. We further extended these allocation rules to mechanisms, \emph{M-ELinUCB-SB}, and \emph{M-SupLinUCB-SB} satisfying EPIC and EPIR properties. We showed our mechanism performs significantly better than the existing mechanism \emph{M-Reg} \cite{Gatti12}. In summary, \emph{M-SupLinUCB-S} is novel,  truthful ConMAB mechanism that outperforms \emph{M-Reg} in every aspect.

Although our mechanisms are randomized, they are game theoretically sound and scalable as compared to \emph{M-Reg}. 
Further, in terms of regret, \emph{M-ELinUCB-SB} and \emph{M-SupLinUCB-S} outperforms \emph{M-Reg} in experiments and theoretically \emph{M-SupLinUCB-S} matches the regret in non-strategic setting up to a factor of $O(n)$ which is the price of truthfulness. 
Though we presented mechanisms for single slot SSA, they can be generalized to multi-slot SSA using similar techniques. Along with SSA, this work can form a baseline for other applications such as crowdsourcing \cite{Biswas15}, smart grids \cite{jain14}, where similar setting arises to learn the stochastic parameters in the presence of strategic agents.  


\bibliographystyle{plain}
\bibliography{main}

\begin{thebibliography}{10}

\bibitem{Abbasi11}
Yasin Abbasi-Yadkori, D\'{a}vid P\'{a}l, and Csaba Szepesv\'{a}ri.
\newblock Improved algorithms for linear stochastic bandits.
\newblock In {\em Proceedings of the 24th International Conference on Neural
  Information Processing Systems}, NIPS'11, pages 2312--2320, USA, 2011. Curran
  Associates Inc.

\bibitem{aggarwal06}
Gagan Aggarwal, Ashish Goel, and Rajeev Motwani.
\newblock Truthful auctions for pricing search keywords.
\newblock In {\em Proceedings of the 7th ACM Conference on Electronic
  Commerce}, EC '06, pages 1--7, New York, NY, USA, 2006. ACM.

\bibitem{arch01}
A.~Archer and \'{E}. Tardos.
\newblock Truthful mechanisms for one-parameter agents.
\newblock In {\em Proceedings of the 42Nd IEEE Symposium on Foundations of
  Computer Science}, FOCS '01, pages 482--, Washington, DC, USA, 2001. IEEE
  Computer Society.

\bibitem{Auer03}
Peter Auer.
\newblock Using confidence bounds for exploitation-exploration trade-offs.
\newblock {\em J. Mach. Learn. Res.}, 3:397--422, March 2003.

\bibitem{auer02}
Peter Auer, Nicolo Cesa-Bianchi, and Paul Fischer.
\newblock Finite-time analysis of the multiarmed bandit problem.
\newblock {\em Machine learning}, 47(2-3):235--256, 2002.

\bibitem{Bab15}
Moshe Babaioff, Robert~D. Kleinberg, and Aleksandrs Slivkins.
\newblock Truthful mechanisms with implicit payment computation.
\newblock {\em J. ACM}, 62(2):10:1--10:37, May 2015.

\bibitem{Bab09}
Moshe Babaioff, Yogeshwer Sharma, and Aleksandrs Slivkins.
\newblock Characterizing truthful multi-armed bandit mechanisms: Extended
  abstract.
\newblock In {\em Proceedings of the 10th ACM Conference on Electronic
  Commerce}, EC '09, pages 79--88, New York, NY, USA, 2009. ACM.

\bibitem{berg2010}
Dirk Bergemann and Juuso V{\"a}lim{\"a}ki.
\newblock The dynamic pivot mechanism.
\newblock {\em Econometrica}, 78(2):771--789, 2010.

\bibitem{Biswas15}
Arpita Biswas, Shweta Jain, Debmalya Mandal, and Y.~Narahari.
\newblock A truthful budget feasible multi-armed bandit mechanism for
  crowdsourcing time critical tasks.
\newblock In {\em Proceedings of the 2015 International Conference on
  Autonomous Agents and Multiagent Systems}, AAMAS '15, pages 1101--1109,
  Richland, SC, 2015. International Foundation for Autonomous Agents and
  Multiagent Systems.

\bibitem{Chu11}
Wei Chu, Lihong Li, Lev Reyzin, and Robert E.~Schapire.
\newblock Contextual bandits with linear payoff functions.
\newblock {\em Journal of Machine Learning Research - Proceedings Track},
  15:208--214, 01 2011.

\bibitem{Nikh09}
Nikhil~R. Devanur and Sham~M. Kakade.
\newblock The price of truthfulness for pay-per-click auctions.
\newblock In {\em Proceedings of the 10th ACM Conference on Electronic
  Commerce}, EC '09, pages 99--106, New York, NY, USA, 2009. ACM.

\bibitem{Gatti12}
Nicola Gatti, Alessandro Lazaric, and Francesco Trov\`{o}.
\newblock A truthful learning mechanism for contextual multi-slot sponsored
  search auctions with externalities.
\newblock In {\em Proceedings of the 13th ACM Conference on Electronic
  Commerce}, EC '12, pages 605--622, New York, NY, USA, 2012. ACM.

\bibitem{iab}
IAB.
\newblock Iab internet advertising revenue report. 2018 first half-year
  results., 2018.

\bibitem{jain18}
Shweta Jain, Sujit Gujar, Satyanath Bhat, Onno Zoeter, and Y~Narahari.
\newblock A quality assuring, cost optimal multi-armed bandit mechanism for
  expertsourcing.
\newblock {\em Artificial Intelligence}, 254:44--63, 2018.

\bibitem{jain14}
Shweta Jain, Balakrishnan Narayanaswamy, and Y.~Narahari.
\newblock A multiarmed bandit incentive mechanism for crowdsourcing demand
  response in smart grids.
\newblock In {\em Proceedings of the Twenty-Eighth AAAI Conference on
  Artificial Intelligence}, AAAI'14, pages 721--727. AAAI Press, 2014.

\bibitem{lai85}
T~Lai.
\newblock Asymptotically efficient adaptive allocation rules.
\newblock {\em Advances in Applied Mathematics}, 6:4--22, 1985.

\bibitem{Lang07}
John Langford and Tong Zhang.
\newblock The epoch-greedy algorithm for contextual multi-armed bandits.
\newblock In {\em Proceedings of the 20th International Conference on Neural
  Information Processing Systems}, NIPS'07, pages 817--824, USA, 2007. Curran
  Associates Inc.

\bibitem{Li10}
Lihong Li, Wei Chu, John Langford, and Robert~E. Schapire.
\newblock A contextual-bandit approach to personalized news article
  recommendation.
\newblock In {\em Proceedings of the 19th International Conference on World
  Wide Web}, WWW '10, pages 661--670, New York, NY, USA, 2010. ACM.

\bibitem{Myr81}
Roger~B Myerson.
\newblock Optimal auction design.
\newblock {\em Mathematics of operations research}, 6(1):58--73, 1981.

\bibitem{Nisan07}
Noam Nisan, Tim Roughgarden, Eva Tardos, and Vijay~V. Vazirani.
\newblock {\em Algorithmic Game Theory}.
\newblock Cambridge University Press, New York, NY, USA, 2007.

\end{thebibliography}
\end{document}